\documentclass[epj,final]{svjour}
\usepackage{latexsym}
\usepackage{graphics}
\begin{document}
\title{Zero-Point Momentum in Complex Media}
\author{B.A. van Tiggelen 
}                     
%
%
\institute{Laboratoire de Physique et Mod\'{e}lisation des Milieux Condens\'{e}s, CNRS /Universit\'{e} Joseph Fourier, BP 166, F-38042
Grenoble Cedex 9, France}
\date{Received: date / Revised version: date}
%
\abstract{
 In this work we apply field regularization techniques to formulate a number of new phenomena related to momentum induced by electromagnetic zero-point fluctuations. We discuss the zero-point momentum associated with magneto-electric media, with moving media, and with 
magneto-chiral media.
} 
\maketitle
\section{Introduction}
\label{intro}

It is well-known and widely accepted that zero-point fluctuations
affect the physics on both microscopic and macroscopic scales,
upon creating forces between materials. The Casimir effect \cite{casi0}- the
attractive force between two macroscopic metallic plates - and
physically equivalent to the Lifshitz effect when it comes to
dielectric media -  is undoubtedly the most famous effect. Also
the $1/r^6$ Van der Waals force and its retarded $1/r^7$
equivalent, the Casimir-Polder force \cite{c-polder}, between microscopic
polarizable atoms can be understood as direct manifestations of
zero-point energy \cite{physicstoday}.

Electromagnetic zero-point energy has been the subject of many fundamental
 work, sometimes heavily
debated in literature, since it often lacks experimental
verification. The most fundamental aspect of vacuum fluctuations
is their Lorentz invariance. It can be shown that \emph{only}
\emph{isotropic radiation with power spectrum} $\omega^3$,
obviously the one associated with  modes whose density is
proportional to $\omega^2/c_0^3$ and whose  zero-point energy is
$\frac{1}{2}\hbar \omega$ \cite{milloni} is Lorentz-invariant. The
\emph{Unruh effect} \cite{unruh} is another fundamental result and concerns
observers with constant acceleration $a$. They see  the zero-point
energy emerge as a Planck law with temperature $kT= \hbar a /2\pi
c_0$ \cite{boyer,milloni}. This effect has never been observed.
On the basis of the relativistic equivalence principle, gravity
should create the same effect, known as Hawking
radiation\cite{hawking}.

The most controversial aspect of vacuum energy is the UV
catastrophe \cite{lkb}. The vacuum energy density is arbitrarily large at
large frequencies. No rigorous mathematical tool seems to exist so
far to deal with this problem. Fortunately, the divergence does
not affect the Casimir force, since it formally drops out, as can
be seen for instance using the Euler-Maclaurin summation formula
\cite{milloni}. Yet, the energy itself may still be an observable.
In 1993, Schwinger \cite{schwinger} attempted to explain the
electromagnetic energy observed from strongly oscillating water
bubbles - an acousto-optical effect called sonoluminescence - in
terms of the zero-point energy released by the contracting bubble.
Upon disregarding vacuum fluctuations with frequencies beyond the
UV, Schwinger concluded that the excess Casimir energy of a sphere
with radius $a$ and dielectric constant $\varepsilon$ be equal to
$E_c \sim \hbar a^3 \omega_c^4/c_0^3 (1-1/\sqrt{\varepsilon})$
which has the right scaling - it decreases with decreasing volume
- and about the right order of magnitude if the cut-off is properly chosen.
Another longstanding
issue, first raised by Dirac in 1934 \cite{dirac34}, is the observation that
zero-point energy should be gravitationally active and should thus
appear as a contribution to the
 cosmological constant in the Einstein equations. If this is true, the divergence of zero-point energy comes in
 rudely. The  Planck length $\sqrt{\hbar G/c_0^3} = 10^{-33} $m
 is the only available cut-off for vacuum modes hand but this would still
 lead to impossible cosmological scenarios. Dirac concluded that the large zero-point energy is absent 
 for a still mysterious reason.

So good motivations exist to search for regularization techniques
that deal with the divergence in a different way. Milton et al
\cite{critic} proposed a more sophisticated regularization scheme,
where only the finite part of the zero-point energy is considered,
due to the finite geometry \cite{barton}. For a spherical bubble
they find $E_c \sim +\hbar(\varepsilon-1)^2/ac_0 $, which has
clearly the wrong scaling to explain sonoluminescence, and which
is orders of magnitude smaller than the estimate by Schwinger. The
same regularizations have been tested for a variety of  space-time
structures to solve the cosmological constant problem  (see
\cite{miltonreview} for a review). Here the sign is also an issue
since the cosmological constant is believed to be positive,
leading to a negative pressure and an expansion that accelerates.
The sign of the Casimir force itself had already been an issue in
the sixties. In 1956 Casimir himself proposed \cite{casi} that
the vacuum force exerted on a spherical metallic shell with
surface $A=4\pi a^2$ might have the similar attractive form $F=-
\alpha \hbar c_0 A/4\pi a^4=- \alpha \hbar c_0/a^2$ as was found
for the plates, though with a different unknown constant
$\alpha>0$. He speculated that this force might stabilize the
Coulomb repulsion $F= +e^2/a^2$ of the electron. This would
provide a first calculation for the fine structure constant, since
stability would impose $e^2/\hbar c_0 = \alpha$. Unfortunately,
the regularized Casimir force on a metallic surface was shown by
Boyer to be repulsive \cite{boyer}.

Only a few years ago, a new controversial effect triggered by
zero-point fluctuations was put forward, this time addressing
their momentum in  magneto-electric (ME) media. In any medium the
dielectric constant can be affected by external electric and
magnetic fields according to $\Delta\varepsilon= \chi
\mathbf{\hat{k}}\cdot (\mathbf{E}_0\times \mathbf{B}_0)$, with
$\mathbf{\hat{k}}$ the unit wave vector. This leads to different
optical properties for photons propagating along or opposite to
the vector $\mathbf{E}_0\times \mathbf{B}_0 \equiv \mathbf{S}_0$
though, unlike the Faraday effect, independent on circular
polarisation. Ref.~\cite{feigel} considered the radiative momentum of photons
in ME media. His predictions are made in a context
that is already controversial in itself, since the momentum of
photons in matter is still heavily debated \cite{brevik}. The
momentum density of zero-point fluctuations in a ME medium  
was found to be,

\begin{equation}\label{feig}
\mathbf{p} = \frac{1}{32\pi^3}(\mu^{-1}+\varepsilon) 
 \frac{\hbar\omega_c^4}{c_0^4} \, \chi\mathbf{S}_0
\end{equation}
Like Schwinger in his attempt to explain sonoluminescence, Feigel
regularized by adopting an UV cut-off for the zero-point spectrum,
arguing that at very high frequencies the ME optical response
should vanish. His choice of a lower cut-off wavelength of $0.1$
nm is based on the fact that optical ME has been observed in the
X-ray. He assumed a typical ME effect $\chi S_0  \approx
10^{-11}$, and with mass densities typically equal to 1 g/cm$^3$
this would lead to typical speeds of $v= 30$ nm/s, likely to be too
small to be measurable. However, our literature study revealed
that ME materials exist such as FeGaO$_3$ for which $\chi S_0
\approx 10^{-4}$ is observed down to wavelengths of order 2 $\AA$ in the X-ray \cite{kubota}. The prediction in
Eq.~(\ref{feig}) would lead to much larger speeds, up to
centimeters per second, that should be observable in experiments.

The calculation of zero-point momentum in ME media puts forward a
revolutionary prediction (by APS Focus \cite{focus} referred to as
``\emph{momentum from nothing}" ), with a clear order of magnitude
estimate. This work is a new occasion to question cut-off
procedures for zero-point modes.  They break the Lorentz-invariance
of the quantum vacuum provocatively and indeed the end-result (1)
is  so much Lorentz-variant, that it is not even likely to be
repairable. But most of all, like in the Schwinger theory of
sonoluminescence and in the cosmological constant debate, the
cut-off procedures give ``inelegant" and ``unreasonable" results.
As for zero-point momentum, the real QED vacuum is known to have a
frequency-independent ME response $\chi \sim \hbar
e^4/m_e^4c_0^7$ \cite{geert}, so that Formula 1 predicts a finite zero-point
momentum density of ``empty" vacuum up to $10^{50}$ times larger
than the momentum density $ \mathbf{E}_0 \times\mathbf{ B}_0/4\pi c_0$
associated with the applied fields. In matter, the cut-off
procedure is often justified as a crude way of dealing with
dispersion, but the results above suggest that this may not be the
whole story, and that in reality the UV catastrophe is nonexistent
for a yet unknown reason.

In this work we investigate how the zero-point momentum emerges if
one applies the field regularization techniques that have been
proposed in literature. This technique would eliminate the
Schwinger theory as an explanation for sonoluminescence
\cite{critic,prlbrevik}. It is not our intention of this work to advocate regularization
techniques. We wish here to come to a quantitative prediction by
\emph{assuming} the validity of these techniques. They are well
defined mathematically and straightforward to implement
numerically, even in symbolic software. However, to our knowledge
nobody has ever been able to assign the removed, diverging terms
to the values of observable constants, as it should be in a good
renormalizable theory. Also experimental tests are rare. Brevik
etal \cite{prlbrevik} and Barton \cite{barton} regularize the zero-point energy of a
dielectric sphere with volume $V= 4\pi a^3/3$ and dielectric
constant $\varepsilon$ and show that this method is equivalent to
a dimensional regularization of the Van der Waals energy between
the atomes constituting the sphere,
\begin{equation}\label{vsw}
     \int_V d^3\mathbf{r} \int_V d^3\mathbf{r}' \,  \left(-\frac{23\alpha^2}{4\pi |\mathbf{r}-\mathbf{r}'|^7} \right)
      \rightarrow + \frac{23(\varepsilon-1)^2}{1536 \pi a}
\end{equation}
In section 2 we show that this regularization appears again in the
expression for zero-point momentum. In a previous Letter we have
already applied the field regularization methods proposed by Kong
and Ravndal \cite{kong} for the ME zero-point momentum in the
Casimir geometry, and concluded that the effect survives the
regularization, but that its value is reduced by some 20 orders of
magnitude. In the present work we address a genuine finite object:
a ME sphere. This makes the  regularized expression above subject
to experimental tests, since the momentum of a finite object is a
measurable quantity, much more than energy. This would be an
indirect test for the regularization of zero-point motion in
general, and such knowledge could be of vital importance to
proceed for instance in the cosmological constant debate. We will
also consider two other situations where zero-point momentum might
show up: a moving sphere and a magneto-chiral sphere.  Both cases
reveal a surprise. For the moving sphere zero-point fluctuations
seem to achieve a (regularized) momentum proportional to the
velocity of the sphere. This would thus contribute to the
\emph{mass} of the sphere! We will show that the problem of a
moving dipole is actually not UV divergent and that a precise
prediction is obtained for the contribution of zero-point modes to
the mass of a polarizable atom. Finally, for a magneto-chiral
sphere, we will present a microscopic argument why the
contribution of zero-point motion to momentum should vanish. We
hope that this gives deeper insight into the microscopic nature of
chirality.

\section{Bi-anisotropic sphere}\label{sec:1}

Starting point of our theoretical study is the set of macroscopic
Maxwell equations - expressed in Gaussian units - applied to bi-anisotropic matter \cite{bi}. Such
media are described by a general linear "constitutive"
relation between the macroscopic electromagnetic fields $\mathbf{D},\mathbf{H%
}$, and the microscopic fields $\mathbf{E},\mathbf{B}$,

\begin{eqnarray*}\label{const}
  \mathbf{D} &=& \mathbf{\varepsilon} \cdot \mathbf{E}   +    \mathbf{\chi}\cdot \mathbf{B} \\
  \mathbf{H} &=& - \chi^{T}\cdot \mathbf{E}  +  \mathbf{\mu}^{-1}\cdot \mathbf{B}
\end{eqnarray*}
The {constitutive} tensors $\varepsilon $ and $\mu $ are assumed
real-valued symmetric, the {constitutive}, bi-anisotropic tensor
$\chi $ is assumed real-valued. In this first work we wish to exclude the
presence of optical dispersion and absorption. In inhomogeneous media all tensors
depend on the position vector $\mathbf{r}$. Time-dependence can be
allowed as well provided the variation is much slower than the
typical cycle oscillation of the electromagnetic fields, so that
we can still work at constant frequency. The best-known case of
optical bi-anisotropy is undoubtedly rotatory power, which can be
described by the symmetric tensor $\chi _{ij}=g\delta _{ij}$, with
$g$ a pseudo scalar, induced by some microscopic chirality. ME
media can be modelled by  the anti-symmetric choice $\chi
_{ij}=\chi (E_{i}^{0}B_{j}^{0}-B_{i}^{0}E_{j}^{0})$, with $\chi $
a scalar. These relations are combined with two Maxwell's
equations applied to harmonic fields $\exp (-i\omega t)$,
\begin{eqnarray}
-i\omega \mathbf{B} &=&+c_0\phi _{\mathbf{p}}\cdot \mathbf{E}
\label{maxwell}
\\
-i\omega \mathbf{D} &=&-c_0\phi _{\mathbf{p}}\cdot \mathbf{H}-4\pi \mathbf{J}%
_{q} \\
i\mathbf{p}\cdot \mathbf{B} &=&0 \\
i\mathbf{p}\cdot \mathbf{D} &=&4\pi \rho _{q}
\end{eqnarray}%
This leads to the following wave equation,

\begin{eqnarray}\label{waveq}
   \left[\frac{\omega^2}{c_0^2} \varepsilon(\mathbf{r})\right.  &-& i\frac{\omega}{c_0}\phi_\mathbf{p} \cdot \chi^T(\mathbf{r}) +
    i\frac{\omega}{c_0}\chi (\mathbf{r})\cdot\phi_\mathbf{p} 
   \nonumber \\ 
   &\, & \left. - \phi_\mathbf{p} \cdot\mu(\mathbf{r})^{-1}\cdot \phi_\mathbf{p}  \right] 
   \cdot \mathbf{E} = -\frac{4\pi i \omega}{c_0^2}  \mathbf{J}_q
\end{eqnarray}
\noindent in terms of the hermitian tensor operator $\phi_{nm,\mathbf{p}} \equiv i\epsilon_{nml}p_l$. From this equation we can identify
the  interaction between matter and radiation,

\begin{eqnarray}\label{inter}
  \mathbf{V}(\mathbf{r},\mathbf{p})\equiv   \frac{\omega^2}{c_0^2} \left[1-\varepsilon(\mathbf{r})\right] &+& i\frac{\omega}{c_0}\phi_\mathbf{p} \cdot \chi^T(\mathbf{r}) - i\frac{\omega}{c_0}\chi (\mathbf{r})\cdot\phi_\mathbf{p}
   \nonumber \\ &- & \phi_\mathbf{p} \cdot\left[1-\mu(\mathbf{r})^{-1}\right]\cdot \phi_\mathbf{p}
\end{eqnarray}
Upon combining the macroscopic Maxwell-equations, the constitutive equations and the Lorentz-force $ \mathbf{E} + \mathbf{v}/c_0 \times \mathbf{B}$, we can arrive at the following
momentum conservation law,

\begin{eqnarray}\label{momcons}
\partial_t \left( \frac{1}{4\pi c_0}\mathbf{E}\times \mathbf{B} + \rho \mathbf{v}\right) = \nabla\cdot \left( -\rho \mathbf{vv} + \mathcal{T}_0\right)
\end{eqnarray}
with the symmetric vacuum stress tensor $\mathcal{T}_0 = (E_iE^*_j+B_iB^*_j)/4\pi - \delta_{ij}\mathcal{E}$ with $\mathcal{E}= (\mathbf{E}\cdot \mathbf{E}^*+ \mathbf{B}\cdot \mathbf{B}^*)/8\pi$ the electromagnetic energy density. Upon integrating Eq.~(\ref{momcons}) far beyond the physical size of the object we get,

\begin{eqnarray}\label{mom}\frac{d }{dt} \int \mathrm{d}^3 \mathbf{r}\, \left( \frac{1}{4\pi c_0}\mathbf{E}\times \mathbf{B}^* + \rho \mathbf{v}\right) = \lim_{r\rightarrow \infty} \oint d\mathbf{S}\cdot \mathcal{T}_0
\end{eqnarray}

To work out the flow of momentum to infinity, expressed by the surface integral on the right, we consider a radiation field $I(\omega,\mathbf{k})$
incident on a conservative, confined bi-anisotropic obstacle. In the far field the electromagnetic polarization is orthogonal to the direction
of propagation $d\mathbf{S}$. As a result the terms $E_iE^*_j$ and $B_iB^*_j$ of $\mathcal{T}_0$ can be seen not to contribute to momentum. After some algebra we find that,

\begin{eqnarray}
\lim_{r\rightarrow \infty} \oint d\mathbf{S}\cdot \mathcal{T}_0
&\sim&  \int_0^\infty \texttt{d}\omega
    \int \textrm{d}\widehat{\mathbf{k}}_\texttt{in} \int \textrm{d}\widehat{\mathbf{k}}_\texttt{out}
  \, I(\omega,\mathbf{k}_\texttt{in})\nonumber \\ &\, & \frac{\textrm{d}\sigma
({\mathbf{k}}_\texttt{in},\mathbf{k}_\texttt{out})}{\textrm{d}\Omega
} \left(  {\hat{\mathbf{k}}}_\texttt{in} -
{\hat{\mathbf{k}}}_\texttt{out}\right)
\end{eqnarray}
Simultaneous P-T symmetry guarantees that $\textrm{d}\sigma (\textbf{k}_\texttt{in},\textbf{k}_\texttt{out}, \mathbf{\chi})  = \textrm{d}\sigma (\textbf{k}_\texttt{out},\textbf{k}_\texttt{in}, \mathbf{\chi}) $ \cite{comment}.
Hence, if the radiation field $I(\omega,\mathbf{k})$ is isotropic, and this is true when the  obstacle is subject to zero-point radiation,
the momentum flow to infinity vanishes. In particular, no inelastic effects occur due to recoil effects.
 As a result, the total momentum

$$ m\mathbf{v} + \int \mathrm{d}^3  \mathbf{r}\, \frac{1}{4\pi c_0}\mathbf{E}\times \mathbf{B}^*$$
is a conserved quantity.  We will refer to the first
term as the kinematic momentum and to the second term as the radiative
momentum. The above formula for total momentum agrees with the more sophisticated
theory by Nelson \cite{nelson} and we refer to this work for a more detailed discussion in relation to the Abraham-Minkowski 
controversy \cite{brevik,loudon} about which term is the real
``radiation momentum" and which part constitutes the genuine
momentum of ``matter".
 In particular, the contribution 
of radiation to momentum found here is  equal to neither the``Abraham value" 
$\frac{1}{4\pi c_0}\mathbf{E}\times \mathbf{H}^* $, nor the ``Minkowski value"
$\frac{1}{4\pi c_0}\mathbf{D}\times \mathbf{B}^*$.  
We emphasize that for an isotropic
monochromatic wave field scattered from a finite object, the space
integral is perfectly finite since the integrand is confined in
and around the object. The problems will appear when integrating
over a power spectrum that diverges itself as $\omega^3$.

Another pertinent remark is that the momentum conservation
expressed by Eq.~(\ref{mom}) continues to be valid if the
constitutive tensors, the tensor $\chi(\mathbf{r})$ in particular,
are time-dependent. Constitutive equations with time-dependent
coefficients can be justified when the variation of the
coefficients is slow compared to the variation of the fields
themselves. This becomes a delicate issue for vacuum fluctuations
that comprise all frequencies and that can thus be arbitrarily
slow. The regularization techniques show that for an object of
size $a$, the typical frequency that contributes to the momentum
equals $c_0/a$. In cut-off procedures even higher frequencies
dominate. This leaves enough room to turn on the external fields
adiabatically. Since a perfect symmetry exists between the wave
vectors $\mathbf{p}$ and $-\mathbf{p}$ if $\chi=0$, we anticipate
$\mathbf{E}\times \mathbf{B}$ to vanish before turning on the
external fields. The conservation law~(\ref{mom}) thus leads us to
the conclusion that after having turned on the fields, the object
achieves a velocity given by,

\begin{eqnarray}\label{mom2}
m \mathbf{v} = - \frac{1}{4\pi c_0} \int \mathrm{d}^3 \mathbf{r}\,
\langle 0 | \mathbf{E}\times \mathbf{B}| 0 \rangle
\end{eqnarray}
 where $\langle 0 | \cdots | 0 \rangle$ stands for vacuum
 expectation. For the product of two electric fields this expectation can be
 obtained from the fluctuation-dissipation theorem, which at zero
 temperature takes the form,
\begin{eqnarray}\label{fd}
 \langle 0 |E_i(\omega,\mathbf{r})&E&^*_j(\omega',\mathbf{r}')| 0 \rangle \nonumber \\  = &-&4\hbar \frac{\omega^2}{c_0^2} \mathrm{Im} \,
 G_{ij}(\omega,\mathbf{r},\mathbf{r}') \times 2\pi \delta(\omega-\omega')
\end{eqnarray}
with $\mathbf{G}$ the (classical) Green's tensor associated with the wave
equation~(\ref{waveq}). It can be straightforwardly  verified that the momentum of zero-point fluctuations
is expressed as,

\begin{eqnarray}\label{momvac}
  {P}_{\mathrm{rad},i}\equiv \frac{1}{4\pi c_0} \int \mathrm{d}^3 \mathbf{r}\, \langle 0 |
\mathbf{E}\times \mathbf{B}| 0 \rangle_i =-\frac{\hbar}{\pi c_0^2} \times \nonumber \\
\mathrm{Im}\int_0^\infty  \mathrm{d}\omega
\int \frac{d^3\mathbf{k}}{(2\pi)^3}\, \omega  \left( k_i
G_{jj}(\omega, \mathbf{k},\mathbf{k}) - k_j
G_{ij}(\omega,\mathbf{k},\mathbf{k}) \right)\nonumber \\
\end{eqnarray}
For a genuine empty vacuum this reduces to the familiar expression $2\times
\int {d^3\mathbf{k}}/{(2\pi)^3} \, \frac{1}{2}\hbar \mathbf{k}$,
which is zero in view of the perfect symmetry in
$\mathbf{k}$.

\bigskip

In the following we shall consider a sphere with a dielectric
constant $\varepsilon$ slightly different from one and a weak
bi-anisotropic tensor $\mathbf{\chi}$, both confined and constant
in the sphere. We shall expand the Green's function in the
potential interaction~(\ref{inter}). Only contributions linear in
$\chi$ can survive the symmetry between $\mathbf{k}$ and
$-\mathbf{k}$. We leave technical details to the Appendices. The
 first order Born approximation to $G$ involves one scattering from the sphere expressed
 by
$\mathbf{G}^{(1)}(\mathbf{k},\mathbf{k})= \mathbf{G}_0(\omega,\mathbf{k})\cdot
\mathbf{V}(\omega,\mathbf{k})\cdot \mathbf{G}_0(\omega,\mathbf{k})$ in terms of the
free-space propagator $\mathbf{G}_0(\omega,\mathbf{k})$. The frequency
integral can easily be performed and we find,

\begin{eqnarray}\label{born1}
  mv_i    \sim   \epsilon_{ijk} \chi_{jk} \, a^3\int \mathrm{d}^3
  \mathbf{k}\, \hbar k
  \end{eqnarray}
Dimensional regularization puts the integral to zero, and no
zero-point momentum - proportional to the volume of the object - is found in this order. Note that this
contribution is  considered by Feigel, and handled using a
cut-off.

\bigskip

We proceed with the second order Born approximation, and collect
the terms proportional to $(\varepsilon-1)\chi$. This involves one normal and one bi-anisotropic scattering. We write  $P_{\mathrm{rad}}^i = \hbar(I_{ijj} - I_{jij})$ with

\begin{eqnarray}\label{born2}
I_{ijl}  = -\frac{1}{\pi c_0^2}\mathrm{Im}\, \int_0^\infty
\mathrm{d}\omega \, \omega \int \frac{d^3\mathbf{k}}{(2\pi)^3}
\frac{d^3\mathbf{k}'}{(2\pi)^3} \, k_i \nonumber \\
\left[\mathbf{G}_0(\mathbf{k}) \cdot
\mathbf{V}(\omega,\mathbf{k},\mathbf{k}')\cdot
\mathbf{G}_0(\mathbf{k}') \cdot
\mathbf{V}(\omega,\mathbf{k}',\mathbf{k})\cdot
\mathbf{G}_0(\mathbf{k}) \right]_{jl} \nonumber \\
\end{eqnarray}
We leave the technical details to Appendix A. Important is that \emph{all} terms are proportional to the
  stocked Casimir-Polder energy, that  we will   regularize  as has been proposed in Eq.~(\ref{vsw}). The final result is,
\begin{equation}\label{momfinal}
    mv_i = -P_{\mathrm{rad}}^i= \eta \frac{\hbar}{a} (\varepsilon-1) \epsilon_{inm}\chi_{mn}
\end{equation}
with (see Appendix) $\eta=(I_0-I_1+C/3-A/3+D/3 -E/2) / 192 \pi^2= 0.007909$. 

\subsection{Magneto-electric sphere }

We can insert the choice for a ME medium \cite{bi,geert1}: $\chi_{nm}=g_{\mathrm{EM}}(E_nB_m-E_mB_n)$ to find that
\begin{equation}\label{momfinalME}
    m\mathbf{v }= 2\eta \times  \frac{\hbar}{a} (\varepsilon-1) g_{\mathrm{EM}} \mathbf{E}\times \mathbf{B}
\end{equation}
We recall that this relation applies after having turned on the fields $\mathbf{E},\, \mathbf{B}$ adiabatically.
For a piece of  FeGaO$_3$ (mass density 4.5 g/cm$^3$, $g_{\mathrm{EM}} {E}{B}\approx 10^{-4}$ ) of size $a=1\mu$m we find the
unmeasurable speed $v= 10^{-20}$ m/s, some 12 orders of magnitude smaller than the value predicted by Eq.~(\ref{feig}).  

\subsection{Moving sphere}
It is well known that a  sphere with a dielectric constant
$\varepsilon$, moving  with a speed $\mathbf{v}$ much smaller than
the speed of light, possesses  a bi-anisotropic tensor $\chi_{ij}=
(1-\varepsilon) \epsilon_{ijk}v_k/c_0$ \cite{milloni}. In this work we
systematically neglect dispersion of the dielectric constant. We
note however that the familiar dispersion for the dielectric
constant beyond the plasma frequency, $\varepsilon=1 -
\omega^2_P/\omega^2$ \cite{jackson} will not be able to render the first Born
approximation~(\ref{born1}) finite. This term would still diverge
like $\int dk \, k$. This suggests that the neglect of dispersion is not at the origin of the UV catastrophe, and that the problem is more fundamental.

Equation~(\ref{momfinal}) thus applies and we can write,
\begin{equation}\label{speed}
   \mathbf{P}_{\mathrm{rad}}= -2\eta \times \frac{\hbar}{a c_0} (\varepsilon-1)^2 \mathbf{v}
\end{equation}
The moving sphere thus drags along with him  a  radiative
zero-point momentum with opposite sign. This is a rather
revolutionary prediction, since it implies that the mass of the
object is reduced by its finite size and its polarizability! We
emphasize that this result follows from a dimensional
regularization that is still arguably controversial. It is
difficult to say whether dispersion would eliminate the
divergence. A scaling argument suggests that a dispersion as
least as fast as $\varepsilon-1\sim 1/\omega^4$ is required to give a
finite result for the second order in the Born expansion.

Note that in this particular case it is possible to bookkeep the diverging momentum of the zero-point fluctuations 
into the bulk mass of sphere, i.e; the one measured for large $a \gg $. For a dielectric sphere of size 1 $\mu$m the change is mass is of order
$10^{-9} m_e$ and thus completely negligible. One could speculate about this effect on atomic scale ($a= a_0$, a polarizability density of order
$a_0^3$ so that $\varepsilon-1 \approx 1$ ). This would yield a mass reduction of order  $10^{-4} m_e$ (or 50 eV), still small enough to be unobserved. It is even more speculative to apply  Eq.~(\ref{speed}) on the  electron scale ($10^{-15}$m). Here the
mass associated with zero-point momentum becomes of the same order of magnitude as the rest mass itself, that is of order $m_e = 0.5 $ MeV. Inspired by the original argument by Casimir to explain electron stability \cite{casi} one could even propose that the momentum of the electron is purely due to zero-point motion. If we propose $\mathbf{P}_{\mathrm{rad}}= \alpha ({\hbar}/{r_0 c_0}) \mathbf{v}:= m_e \mathbf{v}$ we find that $\alpha$ should be equal to the fine structure constant $e^2/\hbar c_0$. Unfortunately, for the dielectric sphere we find the opposite sign, just like the Casimir force was also seen to be repulsive \cite{boyer}, but it is fascinating that this argument gives the right order of magnitude for $\alpha$.
We remark that the value found above for $\eta$ is close to the fine-structure constant.

\subsection{Moving dipole}

In the following we consider a moving electric dipole and
calculate semi-classically the radiative momentum associated with
the zero-point fluctuations, modified by the presence of the
dipole. The polarization is assumed to be point like, and if the
dipole is moving with speed $\mathbf{v}$ the light-matter
interaction becomes

\begin{eqnarray}
\mathbf{V}=   (1-\varepsilon) \frac{{\omega^2}}{c_0^2}U |0\rangle\langle 0 | - i \frac{\omega}{c_0} \mathbf{\phi}_{\mathbf{p}} \cdot \mathbf{\chi}^T U |0\rangle\langle 0 | \nonumber \\ + i\frac{\omega}{c_0} U |0\rangle\langle 0 | \mathbf{\chi} \mathbf{\phi}_{\mathbf{p}} 
\end{eqnarray}
with $U$ interpreted as a small physical volume associated with
the dipole , $\alpha=(\varepsilon-1)$ its polarizability
density, and the bi-anisotropic tensor $\mathbf{\chi} =
(1-\varepsilon) (\mathbf{\epsilon}\cdot \mathbf{v}/c_0) $. The
advantage of this interaction is that the full Born series can be
summed, although momentum integrals have to be regularized
\cite{pr}.  For $\mathbf{v}=0$, the $t$-matrix is found from

\begin{eqnarray}t_0 = -\frac{1}{(\alpha\omega^2)^{-1} + \mathbf{G}_0(\mathbf{r}=0)} =  \frac{-4\pi \Gamma \omega^2/c_0^2}{\omega_0^2 -\omega^2 - \frac{2}{3}i\Gamma \omega_0^2 \omega/c_0}  \nonumber \\
\end{eqnarray}

The second familiar formula is obtained when a momentum
regularization is adopted for the diverging $k$-integral of the
Green's tensor \cite{pr}.  The divergence of the longitudinal part
$\Lambda_L/\omega^2 >0$ can be absorbed into the   polarizability
by defining

$$ \frac{1}{\alpha (0)} := \frac{1}{\alpha} + \Lambda_L$$

A similar ``satisfactory regularization" procedure is not possible
for the transverse Green's tensor. Dimensional regularization
introduces $\Gamma = 1/\Lambda_T$ and the resonant frequency
$\omega_0=c_0(\alpha(0) \Gamma)^{-1/2}$.

The regularization for the isotropic dipole produces an
``acceptable" result for its scattering amplitude. In Appendix B
it is established that the inclusion of bi-anisotropic effects
does not lead to new singularities. We can then use
Eq.~(\ref{momvac}) to calculate the vacuum expectation value for
$\mathbf{E}\times \mathbf{B}$. We find

\begin{eqnarray} \mathbf{P}_{\mathrm{rad}}(\omega) =    \frac{2\hbar }{\pi} \, \mathrm{Im}\, \frac{t_0}{\alpha\omega^2} \  \mathbf{v}
\end{eqnarray}
It can easily be checked that the frequency integral of $\mathrm{Im}\,{t_0}/{\omega^2} $ converges and equals $-(\pi/2) \alpha(0) \omega_0/c_0^2$. Hence we arrive at the final result,

\begin{eqnarray} \mathbf{P}_{\mathrm{rad}} =     - \frac{\alpha(0)}{\alpha}  \frac{\hbar \omega_0}{c_0^2} \mathbf{v}
\end{eqnarray}
We
conclude that the moving dipole drags a momentum associated with
zero point opposite to its kinematic momentum. This can be
interpreted
 as a \emph{reduction }of the kinematic mass. The minus sign was also found earlier for the moving sphere.

 It is surprising
 to see that the ratio of real to \emph{bare} polarizability density  comes in. For a small dielectric sphere difference between $\alpha$ and $\alpha(0)$
 can be attributed to depolarization induced by surface charges,
 but if we want to apply this the model to an atom, $\alpha$  is usually supposed to be unmeasurable.
 We can now imagine two scenarios.
 If this depolarization is negligible, typically true when $\alpha \approx \alpha(0) \approx U$, the front
 factor in $\mathbf{P}_{\mathrm{rad}}$ equals one, and the mass would be reduced by an amount
 $\hbar\omega_0/c_0^2$. For a typical resonant transition at a few eV this would modify the Hydrogen mass by roughly
 one part in $10^9$. This is roughly the same value estimated in the previous section on the basis of a dielectric atom. If however $\alpha \gg \alpha(0) \approx U $, the finite polarizability density is fully governed
 by surface depolarization, described here by the regularization scalar $\Lambda_L$, then  $\mathbf{P}_{\mathrm{rad}} =0$.
 Unfortunately, the present semi-classical approach is not able
 to predict the  value of $\alpha(0)/\alpha$. A  quantum theory is needed.

We expect in general that zero-point motion does not generate
energy flow, not even in bi-anisotropic. This means that the
quantum expectation value of the Poynting vector should vanish. In
a bi-anisotropic the latter is not
 necessarily proportional to the momentum. For the moving dipole it can be  checked explicitly that the
 quantum expectation value of the Poynting vector $\mathbf{S}= c_0 \mathbf{E} \times \mathbf{H}/4\pi $, indeed vanishes.

\section{Magneto-chiral object}\label{section:3}

The optical properties of a homogeneous magneto-chiral (MC) material can be
characterized by a contribution to the index of refraction that is
independent on polarization, and linear in the magnetic field \cite{mc}
Symmetry arguments impose that the sign of this contribution is
different for opposite enantiomeres, and opposite for counter
propagating beams. Thus typically $\Delta n \sim g \mathbf{k\cdot B}_0$ with
$g$ a material pseudo scalar related to microscopic chirality.
This phenomenon can seen as a collective effect of rotatory power,
with optical bi-anisotropy $\chi_{ij} = g\delta_{ij}$, and the
Faraday effect that contributes $iV\epsilon_{ijk}B_k $ to the
dielectric constant, with $V$ the Verdet constant.

If we accept this macroscopic description of optical MC, the
"radiative" momentum of a MC sphere created by zero-point motion
can be calculated in just the same way as was done earlier. The
second order Born approximation generates optical MC by means of
products of chiral and Faraday-type terms, that have to be
regularized when we integrate over all frequencies of the vacuum.
In a attempt to be more realistic one could accept that the Verdet
constant behaves like $V(\omega) = V_0\omega^2 $ up to relatively
large frequencies. On the other hand, rotatory power is expected
to have only little frequency dispersion. Without specifying
details - the calculation is similar to the one for a magneto-electric sphere - we give here the outcome for the momentum obtained by a MC
sphere, after having turned on the magnetic field adiabatically

\begin{equation}\label{wrong}
\mathbf{p}= -0.005098 \, \frac{\hbar V_0 c_0^2g}{a^3}\, \mathbf{B}
\end{equation}

If the assumptions and the regularization proposed above are correct, this
formula would imply that enantiomeres of opposite chirality can be
separated by turning on a magnetic field.

A second approach consists of accepting the unavoidable
heterogeneous structure of space that underlies spatial chirality.
One can propose a simple optical model to describe a chiral
``molecule" in terms of a  chiral distribution of $N>4$ classical
dipoles. If the dipoles are subject to the Zeeman effect, this
molecule exhibits MC properties in the optical scattering, that
are particularly revealed when we average over orientations to
restore spherical symmetry \cite{felipe}.

The scattering amplitude of the MC molecule - linearized in the
external magnetic field - was obtained in Ref.~\cite{felipe}. It
can be inserted into expression~(\ref{momvac}) to find the
radiative momentum. The end result can be expressed in terms of a
trace of a complex $3N \times 3N $ matrix involving two
L\'evi-Civita tensor densities, and the scattering amplitude $t_0$
of the dipoles found earlier.  A straightforward analyses leads us
- quite surprisingly - to  \emph{exactly} the same expression as
 for the ``diffuse
supercurrent" that was considered by us in Ref.~\cite{felipe}. The possibility of such a current, directed along
the magnetic field and not involving the familiar gradient of
energy density (familiar from Fick's law),  was investigated for
  random media with chiral scatterers, but with negative result. In the present context
we thus conclude that the chiral object does not carry any vacuum
momentum, \emph{not at any frequency}, when the magnetic field is turned on.

One can try to analyse this conclusion. In the microscopic
picture, Poynting vector and radiative momentum are proportional
at any point, since locally $\mu=1$ and $\chi=0$.  Since we do not expect any macroscopic energy
current - quantified by the average Poynting vector over some large volume - to
occur in vacuum (yet this statement is hard to prove in
heterogeneous, complex media), we might anticipate that also the 
macroscopic radiative momentum must vanish, 

\begin{eqnarray}
 \mathbf{P}_{\mathrm{rad}} = \int d^3\mathbf{r} \frac{1}{4 \pi c_0} \langle 0|  \mathbf{E} \times \mathbf{B } |0 \rangle = 
\frac{1}{ c_0^2}\int d^3\mathbf{r}  \langle 0 |  \mathbf{S} |0 \rangle =0 \nonumber \\
\end{eqnarray}

This second equality
does not hold in the macroscopic description of MC, and the
two deviate  at any point. In the microscopic picture,
dispersion and spatial structure have been taken into account much
more realistically than in the macroscopic constitutive
description. This example thus shows that one has to be
careful in applying macroscopic Maxwell equations to fundamental
issues whose origin is truly microscopic. The macroscopic,
regularized outcome (\ref{wrong}) is thus probably wrong.

\section{Conclusions}\label{section:4}

The purpose of this work was to come to concrete expressions for the momentum of zero-point motion in complex media.
This constitutes a new and unique occasion to ``test" regularization methods for vacuum properties in experiments, since momentum is 
directly observable, much more than energy.
 Three cases
have been discussed for which the momentum of zero-point motion does not seem to vanish
``trivially". For a non-absorbing sphere subject to both an
external electric and magnetic field we find a radiative momentum
inversely proportional to its radius. To this end regularization
techniques had to be adopted to render the outcome finite. A
confrontation of this prediction to future experiments may thus shed new light
on the validity of regularization methods in general. The same
procedure leads to a radiative momentum of zero-point fluctuations
of a moving sphere. This effect in principle lowers the kinetic
mass of the sphere. The same conclusion is reached for a moving
dipole, thus reassuring that - at least in this case - the
prediction is not an artifact of the macroscopic model. At last,
regularization techniques have been applied to a sphere exhibiting
both rotatory power and the Faraday effect. Here it is possible to
come up with a more microscopic description, using Faraday-active
dipoles in a chiral geometry. In this case the zero-point momentum
is rigorously equal to zero, although the macroscopic models yields a finite result.
 
In the futur we hope to develop fully
quantum-mechanical descriptions of magneto-electric objects and
moving dipoles. The calculations have also been done for idealized media, free from dispersion and absorption. Clearly, this has to be improved
in the future.  The Lorentz-invariance of zero-point motion is also an important
aspect that must be given attention.

\begin{acknowledgement}
The author is indebted to Geert Rikken for many enlightening discussions. This work was supported by the European Advanced Concept Team,
call ACT-RPT-ARIADNA-04-1201.
\end{acknowledgement}

\appendix

\section{Calculation of $I_{ijl}$}

In this Appendix we calculate the $I_{ijl}$ defined by
Eq.~(\ref{born2}). For simplicity we put $c_0=1$. Since the constitutive parameters are assumed
not to vary with frequency, the frequency integral of this object
can be performed using Cauchy contour integration. Since
$\mathbf{G}_0= (1-\mathbf{kk}/\omega^2)(\omega^2-k^2
+i\epsilon)^{-1}$ has a longitudinal part and a part proportional
to the identity, we can essentially discriminate three different
contributions to Eq.~(\ref{born2}), with either 0, 1, or 2
longitudinal propagators in the above expression.

In the absence of any longitudinal part, the frequency integral of
any contribution proportional to $(\varepsilon-1)\chi$ contained
in (\ref{born2}) is of the form
\begin{equation}
\int_0^\infty d\omega\, \omega^4\frac{1}{\left(\omega^2-k^2 + i\epsilon\right)^2}\frac{1}{ \omega^2-k'^2 + i\epsilon} = -\frac{\pi i}{4}
\frac{k+2k'}{(k+k')^2}
\end{equation}
We can write $\mathbf{V}(\omega,\mathbf{k}',\mathbf{k}) =\mathbf{\hat{V}}(\mathbf{k},\mathbf{k}')\theta_{\mathbf{kk}'} $ with
\begin{equation}
\theta_{\mathbf{kk}'} \equiv \int_B \mathrm{d}^3\mathbf{x} \exp[i(\mathbf{k}-\mathbf{k}')\cdot \mathbf{x}]
\nonumber
\end{equation}
where in our case the integral is to be carried out over a sphere. This transforms Eq.~(\ref{born2}) into,
\begin{eqnarray}
I_{ijl}^{(0)} = -\frac{1}{4}(\varepsilon-1) (\epsilon_{jmn}\chi_{lm} +\chi_{jm}\epsilon_{lmn}) \int_B \mathrm{d}^3\mathbf{x} \int_B \mathrm{d}^3\mathbf{y} \nonumber \\
\int \frac{d^3\mathbf{k}}{(2\pi)^3}
\int \frac{d^3\mathbf{k}'}{(2\pi)^3}
 \exp[i\mathbf{k}\cdot (\mathbf{x} -\mathbf{y})]
\exp[-i\mathbf{k}'\cdot (\mathbf{x} -\mathbf{y})]\, \nonumber \\
  \frac{k+2k'}{(k+k')^2}\,  k_i (k_n+k_n') \nonumber
\end{eqnarray}
The four integrals restore complete spherical symmetry, so that
the tensor $k_i(k_n+k_n')$ must lead to a factor proportional to
$\delta_{in}/3$. The factor of proportionality then immediately
follows by contraction. If we re-scale
$k=|\mathbf{x}-\mathbf{y}|p$ and $k'=|\mathbf{x}-\mathbf{y}|q$ we
arrive at the following expression,
 \begin{eqnarray}
 I_{ijl}^{(0)} = -\frac{K(B)}{48 \pi^4} (\varepsilon-1) (I_0 -I_1)(\epsilon_{jmi}\chi_{lm} +\chi_{jm}\epsilon_{lmi})
\end{eqnarray}
where we have introduced the volume integral

$$ K(B)\equiv \int_B \mathrm{d}^3\mathbf{x} \int_B \mathrm{d}^3\mathbf{y}  \frac{1}{|\mathbf{x}-\mathbf{y}|^7}
$$
and the two  scalars

\begin{equation}
I_0 =  \int_0^\infty dp \int_0^\infty dq p^4q^2\frac{p+2q}{(p+q)^2} j_0(p)j_0(q)\nonumber
\end{equation}
\begin{equation}
I_1 =  \int_0^\infty dp \int_0^\infty dq p^3q^3\frac{p+2q}{(p+q)^2} j_1(p)j_1(q)
\nonumber
\end{equation}
A factor $ \mathrm{e}^{-\epsilon(p+q)}$ can be added to ensure convergence.

We can repeat this calculation in the presence of  one longitudinal propagator $(-\mathbf{kk}/\omega^2)(\omega^2-k^2+i\epsilon)^{-1}$.
The frequency integral now becomes,
\begin{equation}
\int_0^\infty d\omega\, \omega^2\frac{1}{\left(\omega^2-k^2 + i\epsilon\right)^2}\frac{1}{ \omega^2-k'^2 + i\epsilon} = \frac{\pi i}{4}
\frac{1}{k(k+k')^2}
\end{equation}
We give the end-result of a long calculation that involves angular averaging of four-rank tensors, that generate Bessel functions of higher order,
and that make the calculation of scalars more involved. We find,
 \begin{eqnarray}
    I_{ijl}^{(1)} &=&  \frac{K(B)}{16\pi^4}  (\varepsilon-1 ) \times \nonumber \\ && \left[
    \left((D_1+\frac{C-A}{15} \right)\chi_{mn} (\delta_{ij}\epsilon_{nlm}+\delta_{il}\epsilon_{njm}) \right. \nonumber \\
    && \left. + \left(D_3+\frac{C-A}{15} \right) (\epsilon_{mli}\chi_{jm}+ \epsilon_{mji}\chi_{lm})\right] 
\end{eqnarray}
with $6D_1+9D_3 =D$ and $12D_1+3D_3=E=E_1+E_2+E_3$
 in terms of
\begin{eqnarray}
A &=&  \int_0^\infty dp \int_0^\infty dq \frac{p^4q^3}{(p+q)^2} j_1(p)j_1(q)
\nonumber\\
C&=&\int_0^\infty dp \int_0^\infty dq \frac{p^5q^2}{(p+q)^2} j_0(p)j_0(q) \nonumber \\
D&=&\int_0^\infty dp \int_0^\infty dq \frac{p^3q^4}{(p+q)^2} j_0(p)j_0(q) \nonumber \\
E&=&\frac{1}{(4\pi)^2}\int  d^3\mathbf{p} \int  d^3\mathbf{q}  \frac{1}{pq} \frac{(\mathbf{p}\cdot \mathbf{q})^2}{p+q} \exp[i(\mathbf{p}-\mathbf{q})\cdot \hat{\mathbf{r}}]\nonumber \\
\end{eqnarray}

The final contribution involves two longitudinal propagators. Since the angular average over a tensor of order 6 appears it is convenient
to perform the contractions first. We find that,
 \begin{equation}
    I_{ijj}^{(2)} -  I_{jij}^{(2)}  = -\frac{K(B) E}{32\pi^4} (\varepsilon-1)\, ) \epsilon_{imn}\chi_{nm}
\end{equation}

We conclude that all contributions generate the stocked
Casimir-Polder energy$K(B)$ of Eq.~(\ref{vsw}) as the only
diverging element.  If we adopt dimensional regularization for this object,
\begin{equation}\label{reno}
    K \rightarrow -\frac{\pi^2 }{12\, a}
\end{equation}

To calculate the  double momentum integrals above we can move the
$p$-integral to  the imaginary axis and  formulate the resulting
integral in the full complex plane using polar coordinates
$(r,\phi)$. We find
\begin{eqnarray}
I_0 &=&  -12\int_0^{\pi/2} d\phi \cos 5\phi \cos \phi \sin^4\phi =0.589\cdots
\nonumber\\
I_1 &=&  -6\int_0^{\pi/2} d\phi \sin^2\phi \cos \phi ( 3\sin 2\phi  \sin 5\phi +2 \cos 3\phi ) \nonumber \\
&=& 4.123\cdots
\nonumber\\
A &=&  4\int_0^{\pi/2} d\phi \sin^2\phi \cos \phi  \cos 2\phi( 2\cos 3\phi  +3 \sin 2\phi \sin 5\phi  ) \nonumber \\
&=&  1.374\cdots
\nonumber\\
C &=&  24\int_0^{\pi/2} d\phi \sin^4\phi \cos \phi  \cos 5\phi \cos 2\phi   = -1.767 \cdots
\nonumber\\
E_1 &=&  -24\int_0^{\pi/2} d\phi \sin^2\phi \cos \phi  (1.5\sin 2\phi \sin 5\phi  +  \cos 3\phi  ) \nonumber \\ &=&  8.246\cdots
\nonumber\\
E_2 &=&  -4\int_0^{\pi/2} d\phi \cos^2\phi   (3+1.5 \sin 2\phi \sin 4\phi \nonumber \\ &+& 3 \sin \phi \sin 3\phi +  \cos \phi \cos 3\phi  ) =  -13.744\cdots
\nonumber\\
E_3 &=&  6\int_0^{\pi/2} d\phi \cos \phi  (6\cos \phi - 2 \cos 2\phi \cos 3\phi  \nonumber \\ &-&\sin^2 2\phi  \cos 5\phi  ) =  24.74\cdots
\nonumber
\end{eqnarray}

\section{Radiative momentum of moving dipole}

The scattering matrix of the moving dipole can be calculated from the Born series ($c_0=1$)

$$\mathbf{T}_{\mathbf{kk}'}  = \mathbf{V}_{\mathbf{kk}'} + \int \frac{d^3\mathbf{k}''}{(2\pi)^3}  \mathbf{V}_{\mathbf{kk}''}\cdot \mathbf{G}_0(\mathbf{k}'')\cdot \mathbf{V}_{\mathbf{k''k}'} + \cdots
$$
which for the bi-anisotropic point dipole can be fully summed up to,
\begin{eqnarray}\label{Tpoint}
\mathbf{T}_{\mathbf{kk}'} = t_0 + \frac{t_0}{(1-\varepsilon)\omega }\left[-i \mathbf{\phi}_{\mathbf{k}} \cdot \mathbf{\chi}^T + i\mathbf{\chi} \mathbf{\phi}_{\mathbf{k}'} \right]
\end{eqnarray}
Here, $t_0$ is the $t$-matrix of the isotropic dipole, given in
the text. We can develop Eq.~(\ref{momvac}) to

\begin{eqnarray}
\mathbf{q}\cdot \mathbf{P}_{\mathrm{rad}} (\omega)
=  -\frac{\hbar }{\pi} \mathrm{Im} \, \mathrm{Tr}\, \int \frac{d^3\mathbf{k}}{(2\pi)^3} \,
\left( \mathbf{k}\cdot \mathbf{q} - \mathbf{k}\mathbf{q} \right) \cdot \mathbf{G}_0(\mathbf{k})\cdot \nonumber \\
\Large\left[(\epsilon\cdot \mathbf{k}) \cdot (\epsilon\cdot \mathbf{v}) + (\epsilon\cdot \mathbf{v})\cdot (\epsilon\cdot \mathbf{k})
\Large\right]
\cdot \mathbf{G}_0(\mathbf{\mathbf{k}}) \nonumber
\end{eqnarray}
 We have  $(\epsilon\cdot \mathbf{k}) \cdot (\epsilon\cdot \mathbf{v}) + (\epsilon\cdot \mathbf{v})\cdot (\epsilon\cdot \mathbf{k})
 =  \mathbf{kv} + \mathbf{vk} -2\mathbf{k\cdot v} $. It is convenient to use the identity, valid to order $\mathbf{v}$,
 $$\mathbf{G}_0(\mathbf{k})\cdot
\Large\left[- \mathbf{kv} - \mathbf{vk} + 2\mathbf{k\cdot v}
\Large\right]
\cdot \mathbf{G}_0(\mathbf{\mathbf{k}}) = \mathbf{G}_0(\mathbf{k}+\frac{1}{2}\mathbf{v}) -\mathbf{G}_0(\mathbf{k}-\frac{1}{2}\mathbf{v})
 $$
 This brings us to
 \begin{eqnarray}
\mathbf{q}\cdot \mathbf{P}_{\mathrm{rad}} (\omega) = \frac{\hbar
}{\pi} \mathrm{Im}\, \mathrm{Tr}\, \int
\frac{d^3\mathbf{k}}{(2\pi)^3} \,
 (\mathbf{q\cdot v} - \mathbf{qv} ) \cdot \mathbf{G}_0(\mathbf{k}) \nonumber
\end{eqnarray}
Since $\int {d^3\mathbf{k}}{(2\pi)^{-3}} \,  \mathbf{G}_0(\mathbf{k}) = -1/\alpha\omega^2 -1/t_0$ this reduces to

\begin{equation}\mathbf{q}\cdot \mathbf{P}_{\mathrm{rad}}(\omega) =  (\mathbf{q\cdot v}) \times \frac{2\hbar }{\pi} \mathrm{Im}\, \frac{t_0(\omega)}{\alpha\omega^2}
\end{equation}

\end{document}